\documentclass[12pt]{article}
\usepackage{graphicx}
\def\bra#1{\left\langle #1\right|}
\def\ket#1{\left| #1\right\rangle}
\newcommand{\bers}{\begin{eqnarray*}}
\newcommand{\eers}{\end{eqnarray*}}
\newcommand{\bt}{\begin{itemize}}
\newcommand{\et}{\end{itemize}}
\def\beq{\begin{equation}}
\def\eeq{\end{equation}}
\def\bea{\begin{eqnarray}}
\def\eea{\end{eqnarray}}
\def\nn{\nonumber}
\def\sss{\scriptscriptstyle}
\def\bd{B_d^0}
\def\bdbar{{\overline{B_d^0}}}
\def\bs{B_s^0}
\def\bsbar{{\overline{B_s^0}}}
\def\barp{{\raise.35ex\hbox
{${\sss (}$}}---{\raise.35ex\hbox{${\sss )}$}}}
\def\bdbarp{\hbox{$B_d$\kern-1.4em\raise1.4ex\hbox{\barp}}}
\def\bsbarp{\hbox{$B_s$\kern-1.4em\raise1.4ex\hbox{\barp}}}
\def\ks{K_{\sss S}}
\def\kbar{{\overline{K^0}}}
\def\roughly#1{\mathrel{\raise.3ex\hbox
{$#1$\kern-.75em\lower1ex\hbox{$\sim$}}}}

\def\mK{m_{\sss K}}
\def\plb#1#2#3{{\it Phys.\ Lett.} {\bf B#1} (#2) #3}

\def\prl#1#2#3{{\it Phys.\ Rev.\ Lett.} {\bf #1} (#2) #3}

\newread\epsffilein 
\newif\ifepsffileok 
\newif\ifepsfbbfound 
\newif\ifepsfverbose 
\newdimen\epsfxsize 
\newdimen\epsfysize 
\newdimen\epsftsize 
\newdimen\epsfrsize 
\newdimen\epsftmp 
\newdimen\pspoints 
\def\epsfbox#1{\global\def\epsfllx{72}\global\def\epsflly{72}%
 \global\def\epsfurx{540}\global\def\epsfury{720}%
 \def\lbracket{[}\def\testit{#1}\ifx\testit\lbracket
 \let\next=\epsfgetlitbb\else\let\next=\epsfnormal\fi\next{#1}}%
\def\epsfgetlitbb#1#2 #3 #4 #5]#6{\epsfgrab #2 #3 #4 #5 .\\%
 \epsfsetgraph{#6}}%
\def\epsfnormal#1{\epsfgetbb{#1}\epsfsetgraph{#1}}%
\def\epsfgetbb#1{%
\openin\epsffilein=#1
\ifeof\epsffilein\errmessage{I couldn't open #1, will ignore it}\else
 {\epsffileoktrue \chardef\other=12
 \def\do##1{\catcode`##1=\other}\dospecials \catcode`\ =10
 \loop
 \read\epsffilein to \epsffileline
 \ifeof\epsffilein\epsffileokfalse\else
 \expandafter\epsfaux\epsffileline:. \\%
 \fi
 \ifepsffileok\repeat
 \ifepsfbbfound\else
 \ifepsfverbose\message{No bounding box comment in #1; using defaults}\fi\fi
 }\closein\epsffilein\fi}%
\def\epsfclipstring{}
\def\epsfsetgraph#1{%
 \epsfrsize=\epsfury\pspoints
 \advance\epsfrsize by-\epsflly\pspoints
 \epsftsize=\epsfurx\pspoints
 \advance\epsftsize by-\epsfllx\pspoints
 \epsfxsize\epsfsize\epsftsize\epsfrsize
 \ifnum\epsfxsize=0 \ifnum\epsfysize=0
 \epsfxsize=\epsftsize \epsfysize=\epsfrsize
 \epsfrsize=0pt
 \else\epsftmp=\epsftsize \divide\epsftmp\epsfrsize
 \epsfxsize=\epsfysize \multiply\epsfxsize\epsftmp
 \multiply\epsftmp\epsfrsize \advance\epsftsize-\epsftmp
 \epsftmp=\epsfysize
 \loop \advance\epsftsize\epsftsize \divide\epsftmp 2
 \ifnum\epsftmp>0
 \ifnum\epsftsize<\epsfrsize\else
 \advance\epsftsize-\epsfrsize \advance\epsfxsize\epsftmp \fi
 \repeat
 \epsfrsize=0pt
 \fi
 \else \ifnum\epsfysize=0
 \epsftmp=\epsfrsize \divide\epsftmp\epsftsize
 \epsfysize=\epsfxsize \multiply\epsfysize\epsftmp
 \multiply\epsftmp\epsftsize \advance\epsfrsize-\epsftmp
 \epsftmp=\epsfxsize
 \loop \advance\epsfrsize\epsfrsize \divide\epsftmp 2
 \ifnum\epsftmp>0
 \ifnum\epsfrsize<\epsftsize\else
 \advance\epsfrsize-\epsftsize \advance\epsfysize\epsftmp \fi
 \repeat
 \epsfrsize=0pt
 \else
 \epsfrsize=\epsfysize
 \fi
 \fi
 \ifepsfverbose\message{#1: width=\the\epsfxsize, height=\the\epsfysize}\fi
 \epsftmp=10\epsfxsize \divide\epsftmp\pspoints
 \vbox to\epsfysize{\vfil\hbox to\epsfxsize{%
 \ifnum\epsfrsize=0\relax
 \includegraphics{#1}%
 \else
 \epsfrsize=10\epsfysize \divide\epsfrsize\pspoints
 \includegraphics{#1}%
 \fi
 \hfil}}%
\global\epsfxsize=0pt\global\epsfysize=0pt}%
 {\catcode`\%=12 \global\let\epsfpercent=
\long\def\epsfaux#1#2:#3\\{\ifx#1\epsfpercent
 \def\testit{#2}\ifx\testit\epsfbblit
 \epsfgrab #3 . . . \\%
 \epsffileokfalse
 \global\epsfbbfoundtrue
 \fi\else\ifx#1\par\else\epsffileokfalse\fi\fi}%
\def\epsfempty{}%
\def\epsfgrab #1 #2 #3 #4 #5\\{%
\global\def\epsfllx{#1}\ifx\epsfllx\epsfempty
 \epsfgrab #2 #3 #4 #5 .\\\else
 \global\def\epsflly{#2}%
 \global\def\epsfurx{#3}\global\def\epsfury{#4}\fi}%
\def\epsfsize#1#2{\epsfxsize}
\pagestyle{plain}

\begin{document}

\begin{flushright}  
UdeM-GPP-TH-01-92\\
YUMS 01-xxx\\
\end{flushright}

\begin{center} 
{\large \bf
\centerline{Using $\bs(t) \to \phi \ks$ to measure $\sin{2\beta}$}}
\vspace*{1.0cm}
{\large Alakabha Datta\footnote{email: datta@lps.umontreal.ca} and David
  London\footnote{email: london@lps.umontreal.ca}} \vskip0.2cm
{\it Laboratoire Ren\'e J.-A. L\'evesque, Universit\'e de
Montr\'eal,} \\
{\it C.P. 6128, succ.\ centre-ville, Montr\'eal, QC, Canada H3C 3J7} \\
\vspace*{0.5cm} {\large C.S. Kim\footnote{email: kim@cskim.yonsei.ac.kr}}
\vskip0.2cm 
{\it Department of Physics, Yonsei University, Seoul 120-749, Korea} \\
\vskip0.3cm

\bigskip
(\today)
\vskip0.5cm
{\Large Abstract\\}
\vskip3truemm
\parbox[t]{\textwidth} {In this talk based on Ref\cite{dattalondon}
 we show that, unlike other pure $b\to d$
  penguin processes, the decay $\bs(t)\to\phi\ks$ is dominated by a
  single amplitude, that of the internal $t$-quark. Thus, the indirect 
CP asymmetry in this
  decay probes $\sin 2\beta$. This cancellation  holds for most part 
of the parameter space and 
  error on $\sin 2\beta$ is less than 10\%. By measuring 
the direct CP asymmetry, one can
  get a better idea of the probable error on $\sin 2\beta$.}

\end{center}
\thispagestyle{empty}
\vskip0.5cm
\setcounter{page}{1}
\baselineskip=14pt

It has been known for many years that the $B$ system is a particularly
good place to test the standard model (SM) explanation of CP
violation. By measuring CP-violating rate asymmetries in the decays of
neutral $B$ mesons to a variety of final states, one can cleanly
extract the CP phases $\alpha$, $\beta$ and $\gamma$ \cite{BCPreview}.
These CP phases are usually obtained from B decays with mostly 
 tree contributions. However penguin contributions are important in many B decays and so
given the importance of such penguin contributions in B decays, one 
is immediately
led to consider CP violation in pure penguin decays. In the
(approximate) Wolfenstein parametrization \cite{Wolfenstein} of the
Cabibbo-Kobayashi-Maskawa (CKM) matrix, there are only two matrix
elements which have a nonzero weak phase: $V_{td} \propto
\exp(-i\beta)$ and $V_{ub} \propto \exp(-i\gamma)$. Thus, assuming
that the penguin amplitudes are dominated by an internal $t$-quark,
one expects that the $b\to s$ penguin amplitude, which involves the
product of CKM matrix elements $V_{tb} V_{ts}^*$, is real, to a good
approximation. Similarly, the weak phase of the $b\to d$ penguin
amplitude ($V_{tb}V_{td}^*$) is $+\beta$. Knowing that the weak phases
of $\bd$-$\bdbar$ and $\bs$-$\bsbar$ mixing are, respectively,
$-\beta$ and 0, this allows us to compute the weak phase probed in
various pure-penguin decay asymmetries \cite{LonPeccei}:
\bea
b \to d & : & {\rm Asym}(\bd(t) \to K^0\kbar) \sim 0 ~, 
\label{Bdbtod} \\
        & & {\rm Asym}(\bs(t) \to \phi \ks) \sim -\sin 2\beta ~, 
\label{Bsbtod} \\
b \to s & : & {\rm Asym}(\bd(t) \to \phi \ks) \sim +\sin 2\beta ~, 
\label{Bdbtos} \\
        & & {\rm Asym}(\bs(t) \to \phi \phi) \sim 0 ~.
\label{Bsbtos} 
\eea
The problem with the above analysis is that the $b\to d$ penguin
amplitude is {\it not} dominated by an internal $t$-quark. In the
quark-level decays $b\to u{\bar u} d$ and $b \to c{\bar c} d$, the
$u{\bar u}$ and $c{\bar c}$ quark pairs can rescatter strongly into an
$s {\bar s}$ quark pair, giving effective $V_{ub} V_{ud}^*$ and
$V_{cb} V_{cd}^*$ contributions to the $b\to d$ penguin decays above.
Buras and Fleischer have estimated that these contributions can be
between 20\% and 50\% of the leading $t$-quark contribution
\cite{ucquark}. And since the $u$- and $c$-quark contributions have a
different weak phase than that of the $t$-quark contribution, this
implies that the weak phase of the $b\to d$ penguin amplitude is {\it
not} $+\beta$, so that the predictions of Eqs.~(\ref{Bdbtod}) and
(\ref{Bsbtod}) are not valid. On the contrary, due to the presence of
these several decay amplitudes, one expects that a weak phase {\it
cannot} be cleanly extracted from the measurement of CP asymmetries in
pure $b\to d$ penguin decays. One also expects to observe direct CP
violation in such decays.

Here, we re-examine the question of the weak phase of the
$b\to d$ penguin for the exclusive decay $\bs(t) \to \phi\ks$
[Eq.~(\ref{Bsbtod})]. As we will show, although the quark-level
contributions from $u$- and $c$-quarks are non-negligible, at the
meson level the matrix elements involving the corresponding $u$- and
$c$-quark operators each vanish, to a good approximation, over a large
region of parameter space. Thus, to the
extent that this cancellation is complete, the CP-violating rate
asymmetry in $\bs(t) \to \phi\ks$ still cleanly probes the weak phase
$\beta$.

We can write the 
amplitude for $\bsbar \to \phi \ks$ as\cite{dattalondon}
\bea
A_s^{\phi\ks} & =& {G_F \over \protect \sqrt{2}}
(V_{ub}V^*_{ud}P_u
+V_{cb}V^*_{cd}P_c +V_{tb}V^*_{td}P_t) ~.
\label{BsphiKsamp}
\eea
where
\beq
P_{u,c} = \bar{c}_6^{u,c} (1-\frac{1}{N_c^2})
\left[\left\langle{O_{LL}}\right\rangle
-2\left\langle{O_{SP}}\right\rangle \right] ~,
\label{Puceqn}
\eeq
where
\bea
\left\langle{O_{LL}}\right\rangle
& = &
\bra{\phi}\bar{s}\gamma_{\mu}(1-\gamma_5)b\ket{B_s}
\bra{\ks}\bar{d}\gamma^{\mu}(1-\gamma_5)s\ket{0} ~, \nonumber\\
\left\langle{O_{SP}}\right\rangle & = &
\bra{\phi}\bar{s}(1-\gamma_5)b\ket{B_s}
\bra{\ks}\bar{d}(1+\gamma_5)s\ket{0} ~.
\label{LLSPdefs}
\eea
(The operator $O_{SP}$ appears due to a Fierz transformation: $(V-A)
\otimes (V+A) = -2 (S-P) \otimes (S+P)$.) On the other hand, the
contribution from the top penguin is more complicated:
\bea
P_t & = &\left[(c_4^t +\frac{c_3^t}{N_c})\left\langle{O_{LL}}\right\rangle
+(c_3^t +\frac{c_4^t}{N_c})\left\langle{O_{LL1}}\right\rangle\right] 
\nonumber\\
& + &
\left[-2(c_6^t +\frac{c_5^t}{N_c})\left\langle{O_{SP}}\right\rangle
+(c_5^t +\frac{c_6^t}{N_c})\left\langle{O_{LR1}}\right\rangle\right]
\nonumber\\
& - &\frac{1}{2}
\left[(c_9^t +\frac{c_{10}^t}{N_c})\left\langle{O_{LL1}}\right\rangle
+(c_{10}^t +\frac{c_9^t}{N_c})\left\langle{O_{LL}}\right\rangle\right] ~,
\eea
where
\bea
\left\langle{O_{LL1}}\right\rangle
& = &
\bra{\phi}\bar{s}\gamma_{\mu}(1-\gamma_5)s\ket{0}
\bra{\ks}\bar{d}\gamma^{\mu}(1-\gamma_5)b\ket{B_s} ~, \nonumber\\ 
\left\langle{O_{LR1}}\right\rangle
& = &
\bra{\phi}\bar{s}\gamma_{\mu}(1+\gamma_5)s\ket{0}
\bra{\ks}\bar{d}\gamma^{\mu}(1-\gamma_5)b\ket{B_s} ~.
\label{tquark1}
\eea
The Wilson's coefficients $c_i$ of the effective Hamiltonian for 
$B$ decays can be found in Ref\cite{tom}.

It is convienient to rewrite $P_{u,c}$ and $P_t$ as
\bea
P_{u,c} & = & \bar{c}_6^{u,c} (1-\frac{1}{N_c^2})X
\left\langle{O_{LL}}\right\rangle ~, \nonumber\\
P_t & = & a_6X\left\langle{O_{LL}}\right\rangle
+ (a_4-a_6 -\frac{1}{2}a_{10})\left\langle{O_{LL}}\right\rangle + 
(a_3+a_5 -\frac{1}{2}a_{9})\left\langle{O_{LL1}}\right\rangle ~,
\label{tquark2}
\eea
where $\left\langle{O_{LL1}}\right\rangle
=\left\langle{O_{LR1}}\right\rangle$,
\beq
a_i = \cases{ c_i + {c_{i-1} \over N_c} ~, & $i = 4,6,10$ ~,\cr
              c_i + {c_{i+1} \over N_c} ~, & $i = 3,5,9$ ~,\cr}
\eeq
and
\beq
X \equiv \left[ 1 - {2 \left\langle{O_{SP}}\right\rangle \over
    \left\langle{O_{LL}}\right\rangle} \right] ~.
\eeq
It is this latter quantity $X$ which is the focus of our attention in
this paper.

Using the fact that
\beq
\langle K_S~(q)|{\bar d}\gamma_{\mu}(1-\gamma_5) s|~0\rangle = i
f_{K_{S}}q_{\mu} ~,
\eeq
along with the equations of motion for the quarks (we assume that $q =
p_d + p_{\bar s}$), it is straightforward to show that
\beq 
X = \left[ 1 - 2 \, {1 \over m_b + m_s} \, {m_{\sss K}^2 \over m_s +
m_d} \right] ~.  
\label{Xdef}
\eeq
However, the key point is the following: taking $\mK = 500$ MeV, $m_b
= 4.9$ GeV, $m_s = 100$ MeV (all at $b$-quark mass scale), and $m_d
\simeq 0$, one finds that $X=0$!  Thus, the matrix elements vanish for
$u$ and $c$ but do {\it not} vanish for $t$. The decay $\bs(t) \to
\phi\ks$ is therefore dominated by a single decay amplitude --- the
$t$-quark penguin contribution --- and a measurement of the
CP-violating rate asymmetry probes the angle $\beta$
[Eq.~(\ref{Bsbtod})].
Note that one can check that, $\bs(t) \to \phi\ks$ is the {\it only} decay involving a
$b\to d$ penguin amplitude for which the ${\cal O}_u$ and ${\cal O}_c$
matrix elements vanish. 

 The measurement of the time-dependent rate $B_s(t) \to
\phi\ks$ allows one to extract both direct and indirect CP-violating
asymmetries. These are defined as follows:
\bea
a^{\sss CP}_{dir} & = & 
{ |A_s^{\phi\ks}|^2 - |{\bar A}_s^{\phi\ks}|^2 \over
 |A_s^{\phi\ks}|^2 + |{\bar A}_s^{\phi\ks}|^2} ~, \nn\\
a^{\sss CP}_{indir} & = & { 
{\rm Im} \left( {A_s^{\phi\ks}}^* {\bar A}_s^{\phi\ks} \right) \over
 |A_s^{\phi\ks}|^2 + |{\bar A}_s^{\phi\ks}|^2} ~,
\eea
where we have taken the weak phase of $\bs$--$\bsbar$ mixing to be
zero. $A_s^{\phi\ks}$ is defined in Eq.~(\ref{BsphiKsamp}), and ${\bar
A}_s^{\phi\ks}$ is obtained from $A_s^{\phi\ks}$ by changing the signs
of the weak phases. Using CKM unitarity to eliminate the
$V_{ub}V^*_{ud}$ term in Eq.~(\ref{BsphiKsamp}), $A_s^{\phi\ks}$ can
be written as
\beq
A_s^{\phi\ks} = {G_F \over \protect \sqrt{2}}
( {\cal P}_{cu} e^{i\delta_c} + 
{\cal P}_{tu} e^{i\delta_t} e^{-i \beta} ) ~,
\eeq
where we have explicitly separated out the strong phases $\delta_c$
and $\delta_t$, as well as the weak phase $\beta$. The magnitudes of
the CKM matrix elements have been absorbed into the definitions of
${\cal P}_{cu}$ and ${\cal P}_{tu}$. Using this expression for
$A_s^{\phi\ks}$, the CP asymmetries take the form
\bea
a^{\sss CP}_{dir} & = & 
{ 2 {\cal P}_{cu} {\cal P}_{tu} \sin\beta \sin\Delta \over
{\cal P}_{tu}^2 + {\cal P}_{cu}^2 + 2{\cal P}_{tu} {\cal P}_{cu}
\cos\beta \cos\Delta} ~, \nn\\
a^{\sss CP}_{indir} & = & 
{ {\cal P}_{tu}^2 \sin 2\beta +
2 {\cal P}_{cu} {\cal P}_{tu} \sin\beta \cos\Delta \over
{\cal P}_{tu}^2 + {\cal P}_{cu}^2 +2 {\cal P}_{tu} {\cal P}_{cu}
\cos\beta \cos\Delta} ~,
\label{CPasymmetries}
\eea
where $\Delta \equiv \delta_t - \delta_c$. From these expressions, we
see that a nonzero value of $X$ corresponds to a nonzero value of
${\cal P}_{cu}$. This in turn leads to a nonzero value of the direct
CP asymmetry $a^{\sss CP}_{dir}$, and also affects the clean
extraction of $\sin 2\beta$ from the indirect CP asymmetry. In order
to compute the error on the measurement of $\sin 2\beta$, we will need
to estimate the size of the ratio ${\cal P}_{cu} / {\cal P}_{tu}$, as
well as the strong phase $\Delta$.

 In our calculation,
we use current quark masses, evaluated at the scale $\mu \sim m_b$.
For the $b$-quark mass, we take $4.35 \le m_b \le 4.95$ GeV. As for
the current strange-quark mass, we vary $m_s(m_b)$ in the
range $0.08 \le m_s \le 0.12$ GeV.

There are also nonfactorizable effects which might give rise to $X\ne
0$. There have have been several attempts to calculate corrections to
the naive factorization assumption. One promising approach is
QCD-improved factorization \cite{Beneke:1999br}, in which one
systematically calculates corrections to naive factorization in an
expansion in $\alpha_s(m_b) \sim 0.2$ and $\Lambda_{QCD}/m_b$. Naive
factorization appears as the leading-order term in this expansion. If
we consider QCD corrections to this term, we note that the $P^{u,c}$
arise already at $O(\alpha_s)$, and so they receive no corrections at
this order. In fact, the $P^{u,c}$ are part of the $O(\alpha_s)$
corrections to the naive factorization results. There are additional
$\alpha_s$ corrections which can be taken into account by the
replacement $a_i \to a_{ieff} =a_i(1+r_i)$ in Eq.~(\ref{tquark2}),
where $r_i \sim O(\alpha_s)$ are process-dependent corrections to the
naive factorization assumption.

In order to test the robustness of the claim that the indirect CP
asymmetry in $\bd(t) \to \phi\ks$ measures $\sin 2\beta$, we perform
the following analysis. We scan the entire parameter space,
calculating the CP asymmetries $a^{\sss CP}_{dir}$ and $a^{\sss
  CP}_{indir}$ of Eq.~(\ref{CPasymmetries}) at each point in this
space. We are especially interested in the quantity $\delta$, which
measures the fractional difference between the indirect CP asymmetry
and the true value of $\sin 2\beta$:
\beq
\delta \equiv { a^{\sss CP}_{indir} - \sin{2\beta} \over \sin{2\beta}
  }~.
\label{deltadef}
\eeq
In particular, we wish to compute what fraction of the parameter space
leads to a given value for $\delta$. This will give us some sense of
the extent to which the asymmetry in $\bd(t) \to \phi\ks$ truly probes
$\beta$. From 
our calculations, we see that $\sin 2\beta$
can be obtained with an error less than 30\% over virtually the entire
parameter space. And this error is reduced to about 10\% in 80\% of
the parameter space. While this should not be interpreted
statistically as some sort of confidence level, it does indicate that
it is quite likely that $\beta$ can be extracted from the indirect CP
asymmetry in $\bs(t)\to\phi\ks$ with a rather small error.

Of course, if the time-dependent rate for $\bs(t)\to\phi\ks$ is
measured, we will have more information than just $a^{\sss
  CP}_{indir}$: we will also measure the direct CP asymmetry $a^{\sss
  CP}_{dir}$. Since $a^{\sss CP}_{dir}$ vanishes if $X=0$, its value
may help us determine the extent to which $a^{\sss CP}_{indir}$ really
measures $\sin 2\beta$. At first glance, the correlation between
$a^{\sss CP}_{dir}$ and $a^{\sss CP}_{indir}$ appears airtight: if
$a^{\sss CP}_{dir}$ is found to vanish, then this must imply that
$X=0$, so that $a^{\sss CP}_{indir}$ yields $\sin 2\beta$.
Unfortunately, things are not quite so straightforward: $a^{\sss
  CP}_{dir}$ is also proportional to the strong phase difference
$\Delta$ [Eq.~(\ref{CPasymmetries})]. Therefore, if $\Delta \simeq 0$,
then $a^{\sss CP}_{dir}$ will vanish even if $X\ne 0$. Thus, this
possibility must be taken into account in evaluating the correlation
between the measurements of $a^{\sss CP}_{indir}$ and $a^{\sss
  CP}_{dir}$.
{}From our calculations, we see that if $a^{\sss CP}_{dir}$ is measured to be
0.1, one can obtain $\sin 2\beta$ from $a^{\sss CP}_{indir}$ with an
error of 5\% (20\%) in $\sim 55$\% ($\sim 95$\%) of the parameter
space. If $a^{\sss CP}_{dir}$ is found to be tiny, then this is
probably due to the fact that $X \simeq 0$, since $\delta < 5$\% over
$\sim 90$\% of the parameter space. However, as discussed above, this
does not hold over the entire space since $a^{\sss CP}_{dir}$ can be
small if $\Delta \simeq 0$, while $X\ne 0$.

Of course, if $X$ does indeed vanish, this may have some negative
practical implications. Specifically, since there are fewer
contributions to the amplitude for $\bs\to\phi\ks$, one might suspect
that the branching ratio will be smaller than that of other pure $b\to
d$ penguin decays. This is indeed the case and we find \cite{dattalondon}
 $ BR[B_s \to \phi K_s] \sim
10^{-7}$, which is very small. However nonfactorizable QCD corrections 
can increase this branching ratio.
Fortunately, the above analysis for $\bs(t)\to\phi\ks$ also applies to
the decay $\bs(t) \to \phi(1680)K_s$, where $\phi(1680)$ is a radially
excited $\phi$. And we can expect the branching ratio $B_s \to
\phi(1680)K_s$ to be almost a factor 10 larger than $B_s \to \phi K_s$
\cite{Datta:2001hd}. This is because the form factor for $B_s \to
\phi$ (or, in general, for any $B \to$ light meson) probes the
high-momentum tail of the $\phi$ wavefunction. As the radially excited
$\phi(1680)$ has more high-momentum components, the form factor for
$B_s \to \phi(1680)$ is enhanced relative to $B_s \to \phi$. On the
other hand, the $\phi(1680)$ decays to $KK^*$, which makes
$\phi(1680)K_s$ more difficult to reconstruct than the $\phi K_s$
final state.

Note that the measurement of CP violation in $\bd(t) \to \phi\ks$
probes $\beta$ [Eq.~(\ref{Bdbtos})]. If the measurement of $\beta$ as
extracted in this mode disagrees with the measurement of $\beta$ from
$\bd(t) \to \Psi\ks$
{\footnote{$\bd(t) \to \Psi\ks$ only measures $\sin{2\beta}$ which allows measurement of $\beta$ up to discrete ambiguities which can be resolved
by measuring $\cos{2 \beta}$ in $ B \to D^{*}\overline{D^{*}}K_s$\cite{tom1}.}
, this will indicate the presence of new physics
in the $b\to s$ penguin amplitude, i.e.\ in the $b\to s$
flavour-changing neutral current (FCNC) \cite{LonSoni}. Similarly, the
value of $\beta$ extracted in $\bs(t)\to\phi\ks$ or $\bs(t) \to
\phi(1680)K_s$ can be compared with that found in $\bd(t) \to
\Psi\ks$. Assuming that the weak phase of $\bs$-$\bsbar$ mixing is
zero --- and this can be tested by looking for CP violation in $\bs(t)
\to \Psi\phi$, for example --- a discrepancy between these two values
points clearly to new physics in the $b\to d$ FCNC. This new physics
might affect $\bd$-$\bdbar$ mixing and/or the $b\to d$ penguin
amplitude. Now, it is quite likely that CP violation in $\bs$ decays
can only be measured at hadron colliders, since one needs an extremely
large boost in order to resolve the rapid $\bs$-$\bsbar$ oscillations.
Since hadron colliders produce copious amounts of $\bd$ and $\bs$
mesons, it should be possible to perform the $\bd(t) \to \phi\ks$ and
$\bs(t) \to \phi\ks$ analyses simultaneously, since the final state is
the same. Thus, by measuring $\beta$ in these decay modes, at hadron
colliders one can test for the presence of new physics in both the
$b\to s$ and $b\to d$ FCNC's.

\bigskip
\noindent
{\bf Acknowledgements}: 
\bigskip 
 This work 
was financially supported by NSERC of Canada.


\end{document}